# Assessment of environmental impacts from authorized discharges of tritiated water from the Fukushima site to coastal and offshore regions

Jakub Kaizer[a,*], Katsumi Hirose[b], Pavel P. Povinec[a]

[a] Faculty of Mathematics, Physics and Informatics, Comenius University, 84248 Bratislava, Slovakia
[b] Laboratory for Environmental Research at Mount Fuji, Okubo, Shinjyuku, Tokyo, Japan
[*] Corresponding author: J. Kaizer (jakub.kaizer@uniba.sk)

## Abstract

In August 2023, the long-planned discharging of radioactive wastewater from the Fukushima Dai-ichi Nuclear Power Plant (FDNPP) started after the confirmation of its feasibility and safety. As this water contains elevated amounts of tritium even after being diluted, a lot of resources have been invested in the monitoring of the Fukushima coastal region where the discharge outlet is located. We compare the first $^3$H surface activity concentrations from these measurements (up to the end of November 2023) with the available background values to evaluate a possible impact of the long-term discharging on humans and environmental levels of the radionuclide of interest in the same or nearby area. From our results, we can conclude that the joint effect of horizontal and vertical mixing has been significant enough to reduce tritium concentrations at the monitored locations in the region close to the FDNPP port two days after the end of the respective phase of the discharging beyond the detection limit of the applied analytical methods (~ 0.3 Bq L$^{-1}$) which is by five orders of magnitude lower than safety limit for drinking water set by the World Health Organization (WHO). Moreover, the distant correlation analysis showed that tritium concentrations at stations located further than 1.4 km were very close to pre-discharge levels (~ 0.4 Bq L$^{-1}$). We also estimated that the $^3$H activity concentration in the offshore Fukushima region would be elevated by 0.01 Bq L$^{-1}$ at maximum over a year of continuous discharging, which is in concordance with the already published modelling papers and much less than the impact of the FDNPP accident in 2011.

## 1. Introduction

The impact of the Fukushima Dai-ichi nuclear power plant (FDNPP) accident on the environment is still under investigation, even though more than ten years have passed since the event (which began on 11 March 2011) when large amounts of radionuclides, including some long-lived ones such as $^{137}$Cs and $^3$H, were released into the atmosphere and hydrosphere (Chino et al., 2011; Tsumune et al., 2012; Povinec et al., 2013, 2021; Aoyama et al., 2016a,b). After the accident, several sporadic releases of contaminated water have been documented in the coastal region offshore Fukushima, mainly due to groundwater flowing from the FDNPP site and runoff caused by heavy rains (Steinhauser et al., 2015; Aoyama, 2018; Sakuma et al., 2022). The groundwater problem has been partially resolved by the construction of two separate barriers. The first is a so-called "ice wall" that consists of vertical underground pipes with a circulating refrigerant (CaCl$_2$ solution) that freezes pore water in adjacent soil layers and consequently reduces their permeability. The wall surrounds reactor Units 1-4 in a way that should decrease of groundwater infiltration into damaged reactors. The second barrier, an impermeable wall on the seaside, has been built on the coast. Its purpose is to prevent potentially contaminated groundwater from flowing into the sea, which has been further supported by the operation of various bypass and drainage systems that allow the pumping out of water from affected aquifers (Shozugawa et al., 2020). While effectively minimizing the amount of radionuclides discharged into the sea, all these types of countermeasures lead to increased volumes of wastewater that must be treated before being disposed of.

In 2013, the FDNPP operator, Tokyo Electric Power Company (TEPCO), finished installing a radioactive water treatment facility called ALPS (Advanced Liquid Processing System). Multi-nuclide removal equipment comprises three distinct separation systems that were designed to extract the most dangerous radionuclides (62 in total) from wastewater to meet the required radiation safety limits. Radionuclide separation is achieved with the use of ferric hydroxide and carbonate co-precipitation methods that can reach an efficiency up to 95% for some species, followed by different sorption techniques based on inorganic ion exchangers, activated carbon, and chelating organic resins (Lehto et al., 2019). In ten years of facility operation, a large amount of ALPS-treated (decontaminated) water has been accumulated in more than 1000 large volume tanks at a FDNPP site, with a total volume of about $1.2 \times 10^6$ m$^3$ (TEPCO, 2023a). Activities of many targeted radionuclides in stored water have decreased below their limit of detection, either as a consequence of the high decontamination factor of the ALPS or physical decay over time if their half-lives were short enough. However, some long-lived radionuclides could not be separated and remained in ALPS-treated water at quantifiable levels (Buesseler, 2020; Bezhenar et al., 2021). This category includes for example $^3$H, $^{14}$C, $^{90}$Sr, $^{99}$Tc, $^{129}$I, and $^{137}$Cs, where tritium seems to be the most problematic one.

Tritium ($^3$H or T) is the radioactive isotope of hydrogen that decays to stable helium while emitting soft beta radiation ($E_{max}$ = 18.6 keV) with a physical half-life of 12.31 years (Haynes, 2017). The dominant form of $^3$H in the environment is tritiated water (HTO), i.e., a water molecule in which an atom of stable hydrogen is replaced by an atom of tritium. Hence, ALPS is not capable of removing this radionuclide from wastewater using common radiochemical procedures. Other possibilities for its separation, such as advanced porous, 2-D layered materials or co-precipitation using heavy water, are only in the development phase and far from being applicable on the industrial scale, as is needed in the case of the FDNPP (Nakamura et al., 2023; Rethinasabapathy et al., 2023). Another option would be electrolytic enrichment that could lead to a significant decrease in the volume of wastewater; however, processing such a large amount of water would require very high costs in terms of energy and time, which makes it unfeasible from the economic point of view (Lehto et al., 2019). The total activity of $^3$H in storage tanks was estimated to be 0.78 PBq (METI, 2021), which is approximately 0.01% of its global inventory.

In addition to the cosmogenic production of $^3$H by interactions of cosmic ray particles with nitrogen and oxygen atoms in the atmosphere, resulting in its worldwide inventory of about 2.2 EBq (UNSCEAR, 2008), tritium was vastly generated during the nuclear weapons testing era when more than 113 EBq entered the environment, which was mostly stored in the oceans. After six decades since the ban on atmospheric tests of nuclear weapons, some 3.7 EBq of bomb-produced tritium is still present in the marine environment (IAEA, 2005). The impact of other anthropogenic sources is at least three orders of magnitude lower and is relevant only on a regional scale; annual releases of $^3$H from all nuclear facilities are estimated to be about 0.1 EBq (UNSCEAR 2000, 2016). The FDNPP, located in the coastal region of the western North Pacific Ocean, was responsible for tritium releases from its accident in 2011 up to 1 PBq (Kaizer et al., 2020). In the case of the same basin, the total controlled discharges of $^3$H from the Tokai nuclear fuel reprocessing facility until 2007 were reported to be roughly 4.5 PBq (Kokubon et al., 2011). Furthermore, the operation of the Rokkasho nuclear fuel reprocessing plant was tested between 2006 and 2008, resulting in releases of about 1.8 PBq of tritium (Anzai et al.,

2008). Although these recent values might seem quite significant, today tritium levels do not pose an immediate and serious threat to living organisms (UNSCEAR, 2016).

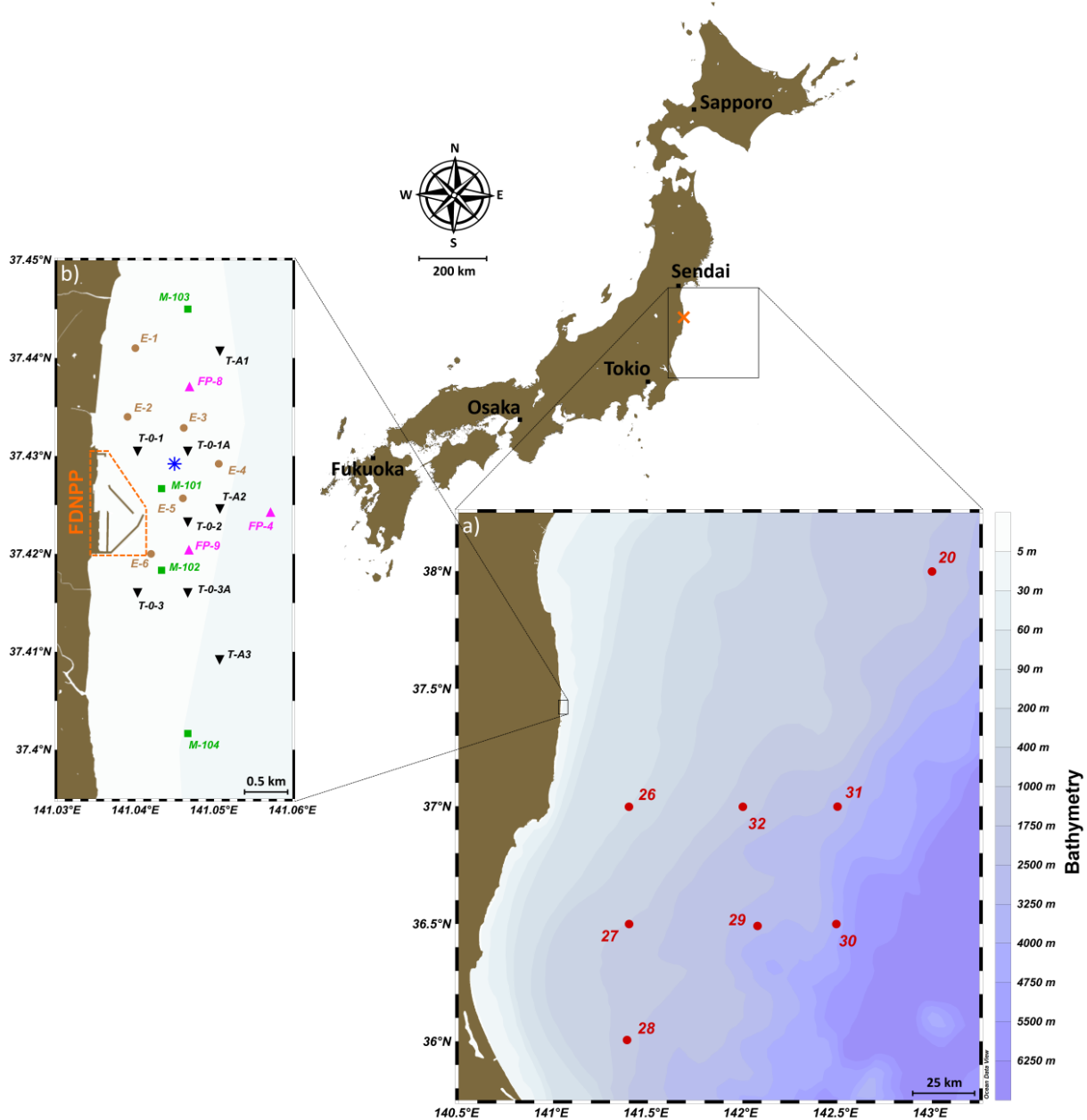

**Fig.1** A map of Japan showing the location of the Fukushima Dai-ichi Nuclear Power Plant (orange cross), (**a**) locations of the stations in the Fukushima offshore region sampled by Povinec et al., 2017 (red circles), (**b**) locations of the discharge outlet (blue asterisk), and stations in the coastal region sampled by the TEPCO (black inverted triangles), Marine Ecology Research Institute (green squares), Fukushima Prefecture (magenta triangle) and Ministry of the Environment (brown circles). The FDNPP harbor area is approximately bordered with an orange dashed line.

Due to the relatively low energy of emitted electrons (5.7 keV on average), tritium is considered radiologically important only in the case of internal exposure, namely ingestion. The intrinsic characteristics of water, i.e., high mobility and turnover, lead to the fact that the biological half-life of tritium in the form of HTO molecules is only about 10 days (Matsumoto et al., 2021). However, part of $^3$H from tritiated water is incorporated into organic matter in the biosphere, forming the so-called organically bound tritium (OBT) fraction, which is much more persistent. Depending on the form, OBT can remain in plant or animal tissues up to a few years (Eyrolle et al., 2018; Ferreira et al., 2023). The metabolic processes responsible for incorporation of $^3$H from tritiated water into organic molecules are ineffective, so that more than 90% of ingested HTO molecules are egested from the human body before being converted into OBT (ICRP, 2002). The low radiotoxicity of tritium is also reflected in regulatory limits for the release or consumption of water. While the limit of the World Health Organization (WHO) for drinking water is set to 10 000 Bq L$^{-1}$, the permissible upper limit for discharge of tritiated water in Japan is 60 000 Bq L$^{-1}$ (WHO, 2017; NRA, 2021; TEPCO, 2023b).

The continuously decreasing available space for the storage of ALPS-treated water at the FDNPP site has forced TEPCO and the Japanese government to look for a solution how to dispose such a large amount of wastewater, which contains significant $^3$H levels which are not uniformly distributed among the tanks; the mean value of $6.5 \times 10^5$ Bq L$^{-1}$ can be deduced from the reported data (METI, 2021; TEPCO, 2023a). In 2021, the authorities decided that ALPS-treated water would be discharged into the coastal region of the western North Pacific Ocean at a rate of 22 TBq y$^{-1}$ and with a maximum activity concentration of 1500 Bq L$^{-1}$, which is achieved by its dilution with seawater (NRA, 2021). The value was calculated so that the radiation dose to the public would be less than $2 \times 10^{-5}$ mSv y$^{-1}$ (METI, 2021). Before the accident, the FDNPP was releasing annually 1.0-2.2 TBq of tritium directly into the marine environment, which means that its discharge rate will increase by one order of magnitude for the next 30 years (Machida et al., 2023). On the other hand, there are several nuclear power plants releasing similar or even higher amounts of tritium in the liquid form in Japan or other nearby countries. For example, in 2023, the Fuqing and Sanmen nuclear power plants in China discharged 121 and 33 TBq of $^3$H (CNNC, 2024a,b), respectively, while it was 36 TBq for the Japanese nuclear power plant in Sendai (KEP, 2024) and 69 TBq in the case of the Kori/Saeul nuclear power plants in the Republic of Korea (KHNP, 2024).

The release of ALPS-treated water has started in late August 2023 after the confirmation of its feasibility and safety by the IAEA, which has been overseeing the whole discharging operation since the beginning of its preparation (IAEA, 2023). An area around the discharge outlet, located approximately 1.5 km from the FDNPP site and at a depth of 12 m, is constantly monitored to check for $^3$H and radiocesium ($^{134,137}$Cs) levels in seawater. In this paper, we compare the first monitoring data provided separately by TEPCO, Marine Ecology Research Institute (MERI), Ministry of Environment and Fukushima Prefecture with the available background values that were determined at different times for the same or nearby region. The aim of such measurements is to evaluate the possible impacts of long-term radionuclide discharges on humans and the marine environment. The presented dataset also includes the results of the screening conducted in 2011 to cover the situation after the FDNPP accident. Furthermore, we calculate how much the released $^3$H concentrations are diluted in seawater with the distance from the point source. The obtained findings are briefly compared with the already published studies, which were based on modeling approaches.

## 2. Data and methods

For the purpose of our investigation, we have compiled five distinctive collections of data that were already published or are available on-line for the general public. The first dataset is taken from the study by Povinec et al. (2017), showing $^3$H activity concentrations in the western North

Pacific Ocean a few months after the FDNPP accident. The screening was carried out as a part of the *Ka'imikai-o-Kanaloa* expedition in June 2011, and included seawater sampled up to 600 km from the FDNPP (from 34 to 38°N, and from 141.5 to 147°E), most of it in the region offshore the Fukushima prefecture. The samples were analyzed using $^3$He in-growth mass spectrometry which guaranteed a low detection limit (about 1 mBq L$^{-1}$) and high precision of the obtained values, typically between 0.2 and 0.5 mBq L$^{-1}$ (Jean-Baptiste et al., 1992; Palcsu et al., 2010). Although more than fifteen stations (st.) were visited at the time of the cruise, we have selected only those that got determined surface $^3$H activity concentrations and that were rather close to the FDNPP and can thus be connected with the current discharging of wastewater. The exact locations of the stations are depicted in Fig. 1a.

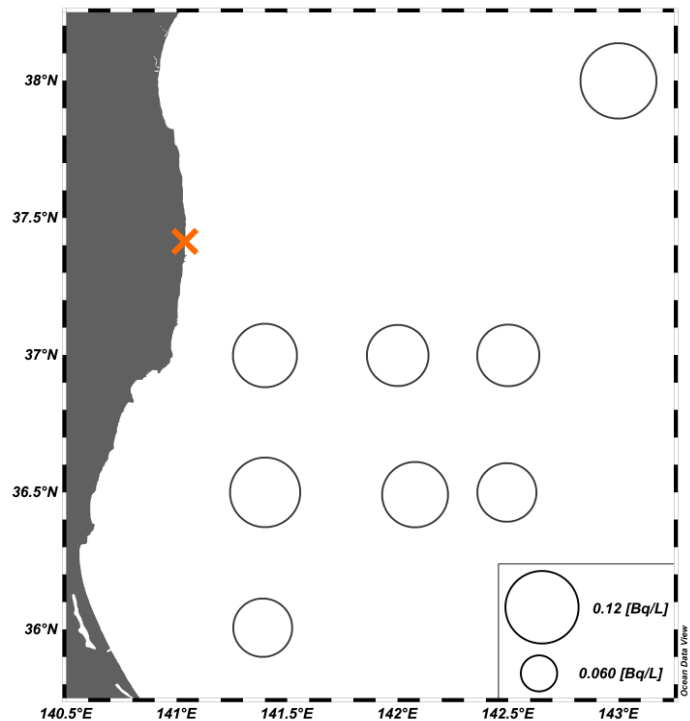

**Fig. 2.** $^3$H activity concentrations in surface seawater in the offshore Fukushima region in June 2011, three months after the Fukushima Dai-ichi Nuclear Power Plant accident, as reported by Povinec et al. (2017). The location of the FDNPP is marked with an orange cross.

The Marine Ecology Research Institute (MERI) was monitoring seawater in the vicinity of the FDNPP from October 2022 to September 2023, as requested by the National Regulation Authority (NRA). Surface samples were collected on a monthly basis, and besides tritium, concentrations of radiocesium and $^{90}$Sr were also reported; the determinations of $^{90}$Sr were provided by a commercial laboratory (NRA, 2023a). While two sampling sites were located less than 300 m from the port of the FDNPP, the other two were established about 2 km north and south of the nuclear power plant site (Fig. 1b). Tritium analyses including electrolytic enrichment were performed using liquid scintillation counting (LSC) with a detection limit of about 0.04 Bq L$^{-1}$. These data represent precise and sensitive measurements of the $^3$H background in seawater, reflecting the situation near the FDNPP prior to and right after the start of $^3$H discharging.

The next set of ³H activity concentration values has been reported by TEPCO which has been constantly monitoring seawater in the coastal area of eastern Japan since the FDNPP accident (TEPCO, 2023c). We focused on 8 sampling stations outside the FDNPP port and around the discharging outlet which is located at approximately 37.43°N and 141.05°E (Fig. 1b). Similarly to the MERI analyses, tritium levels were determined by LSC, though detection limit was generally higher by one order of magnitude (0.3-0.4 Bq L$^{-1}$); in the case of some samples, the improved analytical technique was exploited, leading to a better detection limit (~ 0.07 Bq L$^{-1}$). The regular sampling period has been 7 days; however, samples have been occasionally gathered more frequently. As we were interested in studying the potential influence of the discharging operation, we evaluated only data from the time after its beginning and until the end of its first stage (from August to November 2023).

Several samplings and ³H determinations in the region have also been carried out by the Japanese Ministry of the Environment and by the Fukushima Prefecture (Fukushima Prefecture, 2023; Ministry of the Environment, 2023), using LSC techniques with detection limits of 0.1 and 0.04 Bq L$^{-1}$, respectively. From these data, we selected the samples that were collected at the distance of less than 1.5 km from the discharge outlet (Fig. 1b) and only after the start of the discharging. The Fukushima Prefecture stations located just outside the entrance of the FDNPP harbor and on the coast were not included to minimize their effects on the results of this study.

All seawater samples that were selected for our investigation can be considered surface (depth of 0.5-10 m or less). A complete description of the data we used in the paper can be found in Appendix A as a part of Supplementary data.

3. **Results and discussion**
   *3.1. Potential impact of the ³H discharging in the region offshore Fukushima and its comparison with the FDNNP accident*

To evaluate the potential impact of an early phase of the discharging of the ALPS-treated water on the tritium levels in the surface ocean of the offshore Fukushima region in 2023, we shall first discuss the situation shortly after the FDNPP accident. In June 2011, the ³H activity concentrations in the uppermost layer of seawater in the region (about 47-184 km from the FDNPP) were found between 0.075 and 0.12 Bq L$^{-1}$ (Fig. 2). It is interesting to point out that the highest value was determined for a station that is located furthest from the FDNPP (st. 20). This fact, together with a relatively narrow range of ³H levels in the case of all stations, suggests that Fukushima-derived tritium was distributed uniformly at the time of the measurements. According to the literature (Povinec et al., 2004, 2010, 2017), the pre-Fukushima activity concentration of ³H in the investigated region was estimated to be 0.03-0.06 Bq L$^{-1}$, which means that uncontrolled releases during three months after the FDNPP accident increased this level by a factor of ~ 2. Tritium releases were calculated to be 0.3 ± 0.2 PBq (Povinec et al., 2017), which is at least by one order of magnitude higher than the annual amount currently planned to be discharged. If we assume a similar spreading and dilution rate for both the periods after the FDNPP accident and during the described stage of the discharging, we can expect the area of interest to be affected by < 0.01 Bq L$^{-1}$. Such increase is almost indistinguishable from the background of 0.05-0.06 Bq L$^{-1}$ that can be calculated from results provided by NRA for the offshore region in July and August 2023 (NRA, 2023b), and is in good agreement with the simulation results presented by the TEPCO (TEPCO, 2023d). By utilizing a modelling approach, Zhao et al. (2021) calculated that ³H activity concentration in this particular area would increase up to 0.003-0.01 Bq L$^{-1}$ when as much as 1.2 PBq would be discharged over a single month.

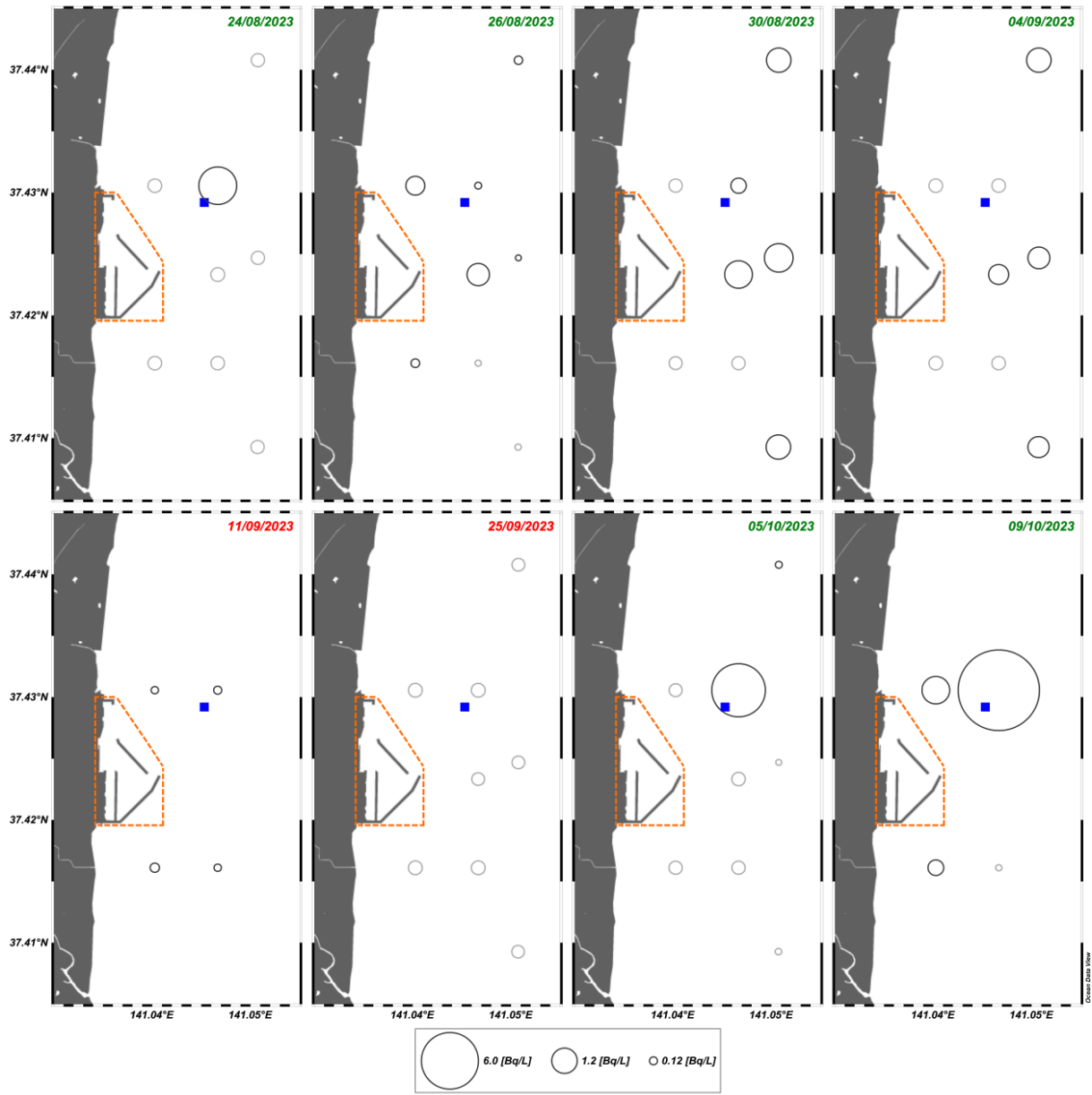

**Fig. 3.** Tritium activity concentrations in surface seawater close to the discharge outlet (blue square) and the port of the Fukushima Dai-ichi Nuclear Power Plant (FDNPP), marked by an orange dashed line, as determined by the TEPCO between 24 August and 9 October 2023 (TEPCO, 2023c). The dates when the discharging was on-going (ceased) are colored in green (red).

Another factor to consider is the influence of the Oyashio Current that flows above the offshore Fukushima region along the 39°N line, making a natural boundary from the northern side. The Oyashio Current is specific in a way that it tends to form a seasonal intrusion that can bring seawater from the subarctic region to the south (down to 36°N), affecting the area of our interest (Tatebe and Yasuda, 2004). Due to the upwelling of deeper waters and the lower rate of deposition of anthropogenic radionuclides from nuclear weapons tests above the 60°N latitude, water masses brought by the intrusion from the subarctic region contain lower amounts of radionuclides (Monetti et al., 1996; Bennett, 2002; Aoyama et al., 2006). In the case of tritium, this impact can be seen in the data from Watanabe et al. (1991), who reported a surface value of 0.12 Bq L$^{-1}$ in the offshore Fukushima region (equivalent to 0.02 Bq L$^{-1}$ after the decay correction to 2011), which was by a factor of 2-3 lower than for the other locations in the western North Pacific Ocean at the time of the expedition; similar effect was also observed for radiocarbon (Tsunogai et al., 1995; Povinec et al., 2017). Therefore, the Oyashio Current and its intrusion represent a potential effect that can enhance dilution of long-term releases of tritiated waters in the investigated region. This effect could be to some extent weakened by the input of $^3$H from the FDNPP port and from rivers in the Fukushima prefecture which was estimated to be cumulatively about 1.4-4.6 TBq per year (Sakuma et al., 2022; Machida et al., 2023).

**Table 1** Tritium activity concentrations in surface seawater from the Fukushima coastal region obtained by the Marine Ecology Research Institute (MERI).

| Station | Date | $^3$H activity concentration [Bq L$^{-1}$] |
|---|---|---|
| **M-101** | 21/04/2023 | 0.12 |
| | 20/05/2023 | 0.058 |
| | 09/06/2023 | 0.082 |
| | 07/07/2023 | 0.11 |
| | 04/08/2023 | 0.070 |
| | 01/09/2023 | 0.066 |
| **M-102** | 21/04/2023 | 0.11 |
| | 20/05/2023 | 0.098 |
| | 09/06/2023 | 0.054 |
| | 07/07/2023 | 0.13 |
| | 04/08/2023 | 0.064 |
| | 01/09/2023 | < 0.052 |
| **M-103** | 21/04/2023 | 0.087 |
| | 20/05/2023 | 0.094 |
| | 09/06/2023 | 0.052 |
| | 07/07/2023 | 0.087 |
| | 04/08/2023 | 0.079 |
| | 01/09/2023 | 0.097 |
| **M-104** | 21/04/2023 | 0.056 |
| | 20/05/2023 | 0.071 |
| | 09/06/2023 | 0.062 |
| | 07/07/2023 | 0.11 |
| | 04/08/2023 | 0.051 |
| | 01/09/2023 | 0.079 |

*3.2. $^3$H activity concentrations in the coastal region and in the vicinity of the FDNPP before and during the discharging in 2023*

The tritium situation in the coastal region near the FDNPP before the start of the discharging can be deduced from the MERI measurements. The results of these analyses, obtained from April to September 2023, are given in Table 1. From this compiled dataset, we can see that the surface $^3$H activity concentration spiked twice (in April and July 2023, reaching values up to 0.13 Bq L$^{-1}$) over the last five months before the beginning of the discharging. These higher values were probably caused by sporadic releases of radioactive waters from the FDNPP port, which has been documented to occur in the past (Aoyama, 2018; Machida et al., 2023). In early August 2023, tritium levels ranged from 0.051 to 0.079 Bq L$^{-1}$. The highest value was determined at st. M-103 (located further to the north), while slightly lower concentrations were found for the station closest to the FDNPP (st. M-101 and st. M-102; Fig. 1b). However, the difference is rather small, and if we realize that the values are only moderately above the declared detection limit (0.04 Bq L$^{-1}$), we can consider them almost the same, especially when we lack knowledge about uncertainties that were not reported by the MERI. Nevertheless, the most important conclusion is that the pre-discharging level of tritium in the coastal Fukushima region was low (< 0.13 Bq L$^{-1}$) and steady.

Temporal changes in surface tritium levels in the vicinity of the FDNPP and at the discharge outlet can be evaluated from the values provided by the TEPCO (Fig. 3 and 4) and from the data obtained by the MERI (Table 1), the Ministry of the Environment (Table 2), and the Fukushima Prefecture (Table 3). Although the latter three data sets are much more limited in comparison to the former data set and show $^3$H activity concentrations only in the first and partially second and third phase of discharging, they are very useful as they were measured with a higher sensitivity, and they can fill several blank spots in the TEPCO set. If we combine them, we can compose a quite sharp picture about the situation after the start of the discharging, in some cases with a resolution of one day.

**Table 2** Tritium activity concentrations in surface seawater from the Fukushima coastal region obtained by the Ministry of the Environment.

| Station | Date | $^3$H activity concentration [Bq L$^{-1}$] |
|---|---|---|
| **E-1** | 25/08/2023 | 0.062 |
|  | 13/10/2023 | 0.76 |
|  | 01/11/2023 | 0.12 |
|  | 14/11/2023 | 3.5 |
| **E-2** | 13/09/2023 | 0.085 |
|  | 14/11/2023 | 0.65 |
| **E-3** | 25/08/2023 | 0.13 |
|  | 14/11/2023 | 0.24 |
| **E-4** | 25/08/2023 | 0.71 |
|  | 13/10/2023 | 0.22 |
|  | 01/11/2023 | 0.13 |
|  | 14/11/2023 | 0.22 |
| **E-5** | 25/08/2023 | 5.0 |
|  | 14/11/2023 | 0.23 |
| **E-6** | 13/09/2023 | 0.097 |
|  | 14/11/2023 | 0.23 |

According to available information (TEPCO, 2024), wastewater with a concentration of $^3$H activity of about 200 Bq L$^{-1}$ was uninterruptedly discharged from 24 August to 10 September. In total, more than 1.2 TBq were released into the sea during this period. The very first tritium data were reported by the TEPCO the same day the releasing had begun when the activity concentration of 2.6 Bq L$^{-1}$ was determined at st. T-0-1A, located very close to the discharge outlet (Fig. 3). As all other stations in the monitoring network showed values below the detection limit (< 0.34 Bq L$^{-1}$), it is difficult to postulate how significant the impact was in the rest of the area. On 25 August, tritium levels were found to be 0.71 and 5.0 Bq L$^{-1}$ at st. E-4 and st. E-5, respectively, while st. E-2 and st. E-3 were almost unaffected (Table 2), suggesting that the main part of the discharged radioactive water was spreading to the south. This conclusion is supported by the TEPCO data from the following day when the highest $^3$H concentration of 0.92 Bq L$^{-1}$ was measured at st. T-0-2, while stations located more to the south (T-0-3A and T-A3) showed values below the detection limit (< 0.072 Bq L$^{-1}$). Increased activity (0.66 Bq L$^{-1}$) was also observed at st. T-0-1 that had been established east of the discharge outlet. The influence of the discharging was amplified on 30 August, reaching $^3$H concentrations up to 1.1-1.5 Bq L$^{-1}$ in the southern locations and at the northernmost station (T-A1) where 1.1 Bq L$^{-1}$ was determined (Fig. 3).

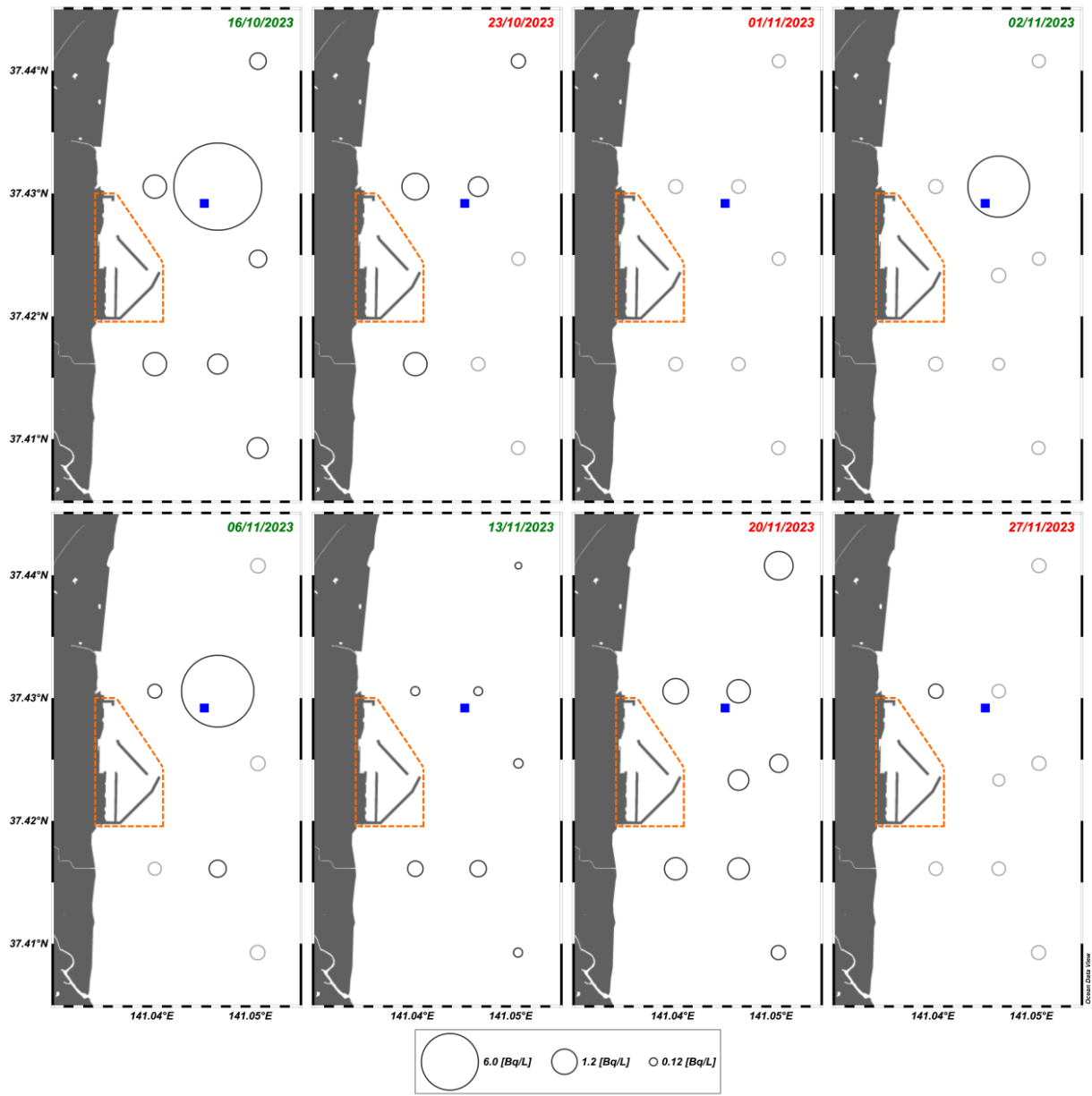

**Fig. 4.** Tritium activity concentrations in surface seawater close to the discharge outlet (blue square) and port of the Fukushima Dai-ichi Nuclear Power Plant (FDNPP), marked by an orange dashed line, as determined by the TEPCO between 16 October and 27 November 2023 (TEPCO, 2023c). The dates when the discharging was on-going (ceased) are colored in green (red).

Despite the higher values at the end of August mentioned in the previous paragraph, tritium concentrations were comparable to the pre-discharge levels at the MERI stations, which can be seen from the values on 1 September (Table 1). With respect to the position of these stations, it would imply that tritium was transported mostly in the eastern/southeastern direction. The same outcome can be deduced from information gathered from stations that were screened by the Fukushima Prefecture two days later when the $^3$H activity concentrations at st. FP-4 and st. FP-

9 were found to be by a factor of ~ 2-3 higher than at st. FP-8 (Table 3). It can be assumed that a similar distribution pattern continued in the following days as $^3$H concentrations at the TEPCO stations did not change significantly between 30 August and 4 September, though the values were lower by about 50-70% in the case of the latter analyses (Fig. 3). The only exception was st. T-A1, where the $^3$H level remained the same during this time.

On 11 September, one day after the first phase of the discharge operation had been ceased, the $^3$H activity concentration was in the range 0.10-0.16 Bq L$^{-1}$ (Fig. 3). Unfortunately, only four TEPCO stations had been sampled, so the situation in the northern and middle part of the region was not known. Due to the lack of the input of "fresh" tritium and continued dilution, its concentrations decreased to even lower values, which is documented by measurements at st. E-2 and st. E-6, yielding 0.085 and 0.097 Bq L$^{-1}$, respectively (Table 2). During the time of preparation of a next phase (approximately three weeks), the monitoring was carried out by a less sensitive LSC method, resulting in the fact that only detection limits (0.29-0.35 Bq L$^{-1}$) were obtained at all TEPCO stations; an example from this period (25 September) is depicted in Fig. 3.

**Table 3** Tritium activity concentrations in surface seawater from the Fukushima coastal region obtained by the Fukushima Prefecture.

| Station | Date | $^3$H activity concentration [Bq L$^{-1}$] |
|---|---|---|
| **FP-4** | 03/09/2023 | 0.15 |
|  | 12/10/2023 | 0.05 |
|  | 09/11/2023 | 0.17 |
| **FP-8** | 03/09/2023 | 0.08 |
|  | 12/10/2023 | 0.05 |
|  | 09/11/2023 | 0.28 |
| **FP-9** | 03/09/2023 | 0.12 |
|  | 12/10/2023 | 0.27 |
|  | 09/11/2023 | 1.6 |

The second phase of the discharging started on 5 October and lasted for 18 days. The average $^3$H activity concentration of discharged ALPS-treated water after dilution was calculated to be about 200 Bq L$^{-1}$ while a total amount of about 1.3 TBq was cumulatively released into the ocean during this phase (TEPCO, 2024). The impact was clearly visible right from the beginning when a high concentration of $^3$H (5.2 Bq L$^{-1}$) was determined at st. T-0-1A (Fig. 3). All other TEPCO stations seemed to be unaffected (st. T-A1) or the effect was below the measuring capabilities of the applied analytical procedure. On 9 October, only samples from half of the locations were analyzed; the concentration at the station closest to the outlet (st. T-0-1A) increased by a factor of more than 2, reaching the value of ~ 12 Bq L$^{-1}$. Tritium was also partially dispersed toward the coast of Fukushima which can be deduced from the values obtained at st. T-0-1 (1.4 Bq L$^{-1}$) and st. T-0-3 (0.45 Bq L$^{-1}$). Partial southward spreading can be concluded from samples collected three days later, as the $^3$H activity concentration at st. FP-9 was determined to be 0.27 Bq L$^{-1}$ while the other two stations showed values equivalent to the background level of 0.05 Bq L$^{-1}$ (Table 3). However, the impact on the coast was still visible next day when 0.76 Bq L$^{-1}$ was observed at st. E-1 while it was much lower at st. E-4 (Table 2). The $^3$H concentration remained very high at st. T-0-1A on 16 October, peaking at ~ 14 Bq L$^{-1}$, which would suggest that the dispersion conditions did not change dramatically between the TEPCO measurements, and that the influence of vertical motion of seawater was less important than during the first phase (Fig. 4). If we compare the data, we can see that released tritium was evenly transported and distributed in the area, impacting a rest of the stations to the level in the range of 0.5-1.0 Bq L$^{-1}$. No more monitoring data were obtained until 23 October when the locations were sampled one day after the end of the second phase of the discharging. In less than twenty hours, tritium activity concentrations decreased by a factor of about 20 at st. T-0-1A while they were below the detection limit (< 0.33 Bq L$^{-1}$) at the southeastern stations. On the other hand, same (st. T-0-3) or even slightly higher (st. T-0-1) values were determined at the near coast stations, meaning that observable dilution processes were still ongoing at that time. The vast majority of tritium in the surface layer was dispersed during the following week, as its concentrations were < 0.31-0.35 Bq L$^{-1}$ at the TEPCO stations, and 0.12-0.13 Bq L$^{-1}$ at st. E-1 and st. E-4 (Table 2).

The discharging operation was resumed on 2 November, with a tritium activity of roughly 1.1 TBq introduced into coastal seawater until 19 November. The $^3$H activity concentration of wastewater was estimated to be ~ 190 Bq L$^{-1}$ (TEPCO, 2024). The same as in the case of the second phase, the level of tritium above the detection limits was observed only at st. T-0-1A (6.9 Bq L$^{-1}$), located less than 200 meters north of the discharge outlet (Fig. 4). Since only two hours had passed between the start of the discharging and the actual sampling, we can assume that the delay was too short for released tritium to affect more distant stations. During the following week, a similarity to the distribution and transport pattern of the second phase persisted; the $^3$H activity concentration at st. T-0-1A increased to 9.5 Bq L$^{-1}$ and remained unmeasurable at the other locations, expect for st. T-0-3A where it was slightly above the detection limit (0.54 Bq L$^{-1}$). From the data obtained on 9 November by the Fukushima Prefecture (Table 3), we can see that a major part of the tritium activity was detected south from the outlet at st. FP-9 (1.6 Bq L$^{-1}$), while two other stations to the north and east were affected only slightly (0.17-0.28 Bq L$^{-1}$). The dispersion in the southern direction was diminished almost completely on 13 November when st. T-0-3A, T-0-3 and T-A3 showed values of 0.44, 0.49 and 0.15 Bq L$^{-1}$, respectively, while $^3$H levels at the northern stations were in the range of 0.082-0.16 Bq L$^{-1}$. On 14 November, ALPS-treated water was probably streaming from the discharge point mainly to the north-west which can be assumed from high $^3$H concentrations reported for st. E-2 (0.65 Bq L$^{-1}$) and st. E-1 (3.5 Bq L$^{-1}$), and from much lower values (0.22-0.23 Bq L$^{-1}$) reported for stations in other directions (Table 2). This substantial spreading of tritium did not last long, as relatively high surface concentrations (0.60-1.5 Bq L$^{-1}$) were observed at almost all TEPCO stations on 20 November, one day after the end of the discharging. In the next seven days, the values decreased very close to or below the detection limits (0.26-0.38 Bq L$^{-1}$).

### 3.3. Spatial variation of $^3$H levels near the FDNPP during the discharging in 2023

Besides the evaluation of temporal changes, we analyzed quite extensive data set from TEPCO to investigate a presumable correlation between the average surface $^3$H activity concentration at the respective station and its distance from the outlet; the results are depicted in Fig. 5. The average tritium levels were calculated for the whole discharging period (from 24 August to 27 November) not only from measured values but also from detection limits; if we had excluded detections limits, the obtained values for some stations would have been overestimated. The activity concentration at the outlet was equal to the mean value for the discharged diluted ALPS-treated (200 Bq L$^{-1}$). The obtained correlation was divided into three distinct sections (i)-(iii) and each one of these was fitted separately using a linear fitting function to better comprehend a change in the decrease rate of the tritium activity concentration with distance.

As it is clear from Fig. 5, the highest average $^3$H level during the discharging periods (3.4 Bq L$^{-1}$) was determined for st. T-0-1A, located closest to the outlet, which is in agreement with the earlier results. In less than 200 meters, the initial (outlet) value decreased by a factor of 60 with a rate of 1.0 Bq $L^{-1}$ per meter. Tritium concentrations at three further locations (T-0-1, T-0-2, T-A2) were in the lower range of 0.43-0.56 Bq $L^{-1}$, meaning that the rapid rate decreased to 0.0056 Bq $L^{-1}$ $m^{-1}$ within some 0.7 km. Such significant change in the rate implies that the dynamic character of the movement of water masses played an important role in the distribution of released tritium in the region. Between 0.7 and 1.5 km from the outlet, the activity concentration of $^3H$ declined only slightly, reaching the range of 0.36-0.49 Bq $L^{-1}$. The most distant station (T-A3), located about 2.3 km from the discharge outlet, showed a similar average value of 0.41 Bq $L^{-1}$, however, 9 out of 14 measurements used in calculations were only detection limits. These findings suggest that the current data were determined with insufficient sensitivity, thus they do not allow us to precisely evaluate the dilution effect beyond 1.5 km. On the other hand, we can safely conclude from the presented results that the impact of the discharging on $^3H$ levels in the distance further than 2.5 km would be less than 0.4 Bq $L^{-1}$. A similar outcome was suggested in a recent paper by Maderich et al. (2024) who exploited a box model POSEIDON to demonstrate that an area of 10 $km^2$ around the FDNPP would be uniformly influenced on the level of 0.3-0.4 Bq $L^{-1}$ if the overall discharging would last for 30 years. Additionally, diffusion simulations based on the meteorological and oceanographic conditions between 2014 and 2020 suggested annual average $^3H$ activity concentration of 0.12 Bq $L^{-1}$ for the very same area, assuming a total discharge of 22 TBq at an even pace throughout the year (TEPCO, 2023d).

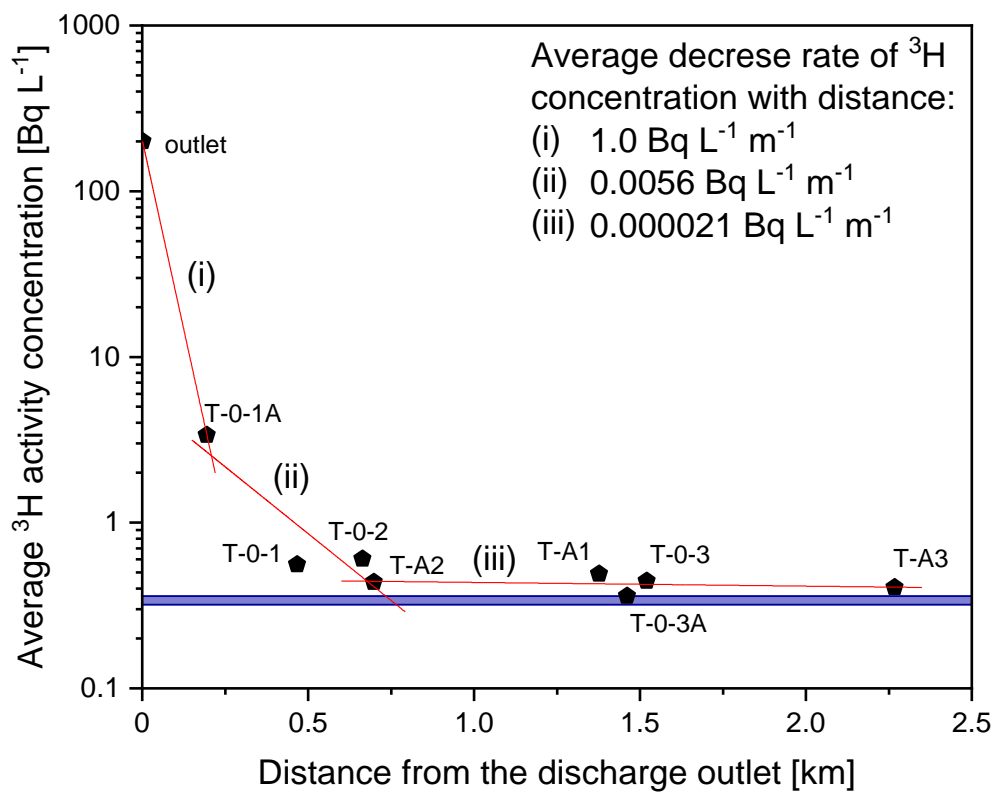

**Fig. 5.** A correlation between the average tritium level at a respective TEPCO station and the distance from the discharge outlet (note that the Y-axis is given in the logarithmic scale). Each subset of the data is fitted by a separate linear function (i)-(iii) to calculate an average decrease rate. Tritium levels observed at distances greater than 1.2 km from the outlet are close to the pre-discharge values (< 0.32-0.36 Bq $L^{-1}$) which were obtained at the stations on 21 August 2023 (royal blue line).

In the distance greater than 1.2 km, tritium activity concentrations were found to be close to the pre-discharge values (< 0.32-0.36 Bq $L^{-1}$, possibly affected by reporting only detection limits) which were obtained at the TEPCO stations on 21 August 2023 (Fig. 5). It is possible that the released ALPS-treated water was moving from the outlet in the form of a narrow plume that was rapidly changing its direction, which would explain why tritium levels at the monitoring stations decreased so greatly within a quite short distance. The spatial and temporal density of the current monitoring network may be simply insufficient to spot such high variability. Therefore, increasing the number of stations and frequency of collecting of surface and subsurface samples during future discharging periods could lead to even more interesting findings.

## 4. Conclusions

The large-scale monitoring provided by the TEPCO and other involved authorities has brought us a good opportunity to investigate a fate of tritium released during the very first stage of the discharging of wastewater from the Fukushima Dai-ichi Nuclear Power Plant (FDNPP) (up to the end of November 2023), planned to continue over the next decades. We can conclude from our results that the dispersion character of the discharged tritium in the sea region close to the FDNPP port is quite dynamic, since the distribution of surface $^3$H was changing by orders of magnitude, sometimes within a few days. The joint effect of horizontal and vertical mixing was significant enough to reduce surface activity concentrations at the monitored locations two days after the end of the respective phase of the discharging beyond the detection limit of applied analytical methods (~ 0.3 Bq $L^{-1}$) which is much lower than the WHO limit for drinking water (10 000 Bq $L^{-1}$). Furthermore, the distant correlation analysis showed that stations located further than 1.4 km from the outlet were impacted by the discharging to a similar level while the highest average $^3$H activity concentration was found at a site closest to the outlet. Because of the observed high dilution and short biological half-life of $^3$H, we can assume that accumulation of discharged tritium in the biomass, e.g., in the form of organically bound tritium (OBT), will be practically negligible. On the other hand, if we consider that the discharging will last for years, an increase in OBT could be eventually detected with the use of advanced analytical techniques.

Although radiological aspects are the most critical when evaluating such operation as the discharging of radioactive water, its influence on environmental levels of long-lived radionuclides from the oceanography perspective may also be important. We estimated that $^3$H activity concentration in offshore waters of the Fukushima region (about 47-184 km from the coastline) would be elevated by 0.01 Bq $L^{-1}$ at maximum over one year of continuous discharging, which is in concordance with the already published modeling papers and much lower than the impact of the FDNPP accident in 2011. To distinguish such a low increase from the background given by a pre-discharge value of 0.05-0.06 Bq $L^{-1}$ (NRA, 2023b), or pre-FDNNP accident value of 0.03-0.06 Bq $L^{-1}$ (Povinec et al., 2004, 2010, 2017) it will be necessary to utilize very sensitive methods $^3$H analysis, e.g., $^3$He in-growth mass spectrometry with a detection limit of about 1 mBq $L^{-1}$. Even though the expected radiological impact on the marine environment would be negligible, the potential benefits may be worth the effort, as tritium can be used as an excellent tracer for studying the pathways and circulation of water masses in the northwestern Pacific Ocean.

**Acknowledgments**

A partial support provided by the EU Operational Program Integrated Infrastructure for the project Advancing University Capacity and Competence in Research, Development and Innovation (ACCORD) (ITMS2014, no. 313021X329), and by the Slovak Science and Grant Agency (VEGA-1/0625/21) is highly acknowledged.

**References**

Anzai, K., Keta, S., Kano, M., Ishihara, N., Moriyama, T., Okamura, Y., Ogaki, K., Noda, K., 2008. Radioactive effluent releases from Rokkasho reprocessing plant (2) – liquid effluent. 12. In: International Congress of the International Radiation Protection Association (IRPA), Buenos Aires, Argentina, 19-24 October 2008.

Aoyama, M., 2018. Long-term behavior of $^{137}$Cs and $^{3}$H activities from TEPCO Fukushima NPP1 accident in the coastal region off Fukushima, Japan. J. Radioanal. Nucl. Chem. 316, 1243–1252.

Aoyama, M., Hamajima, Y., Hult, M., Uematsu, M., Oka, E., Tsumune, D., Kumamoto, Y., 2016a. $^{134}$Cs and $^{137}$Cs in the North Pacific Ocean derived from the March 2011 TEPCO Fukushima Dai-ichi Nuclear Power Plant accident, Japan. Part one: surface pathway and vertical distributions. J. Oceanogr. 72, 53–65.

Aoyama, M., Hirose, K., Igarashi, Y., 2006. Re-construction and updating our understanding on the global weapons tests $^{137}$Cs fallout. J. Environ. Monit. 8, 431.

Aoyama, M., Kajino, M., Tanaka, T.Y., Sekiyama, T.T., Tsumune, D., Tsubono, T., Hamajima, Y., Inomata, Y., Gamo, T., 2016b. $^{134}$Cs and $^{137}$Cs in the North Pacific Ocean derived from the March 2011 TEPCO Fukushima Dai-ichi Nuclear Power Plant accident, Japan. Part two: estimation of $^{134}$Cs and $^{137}$Cs inventories in the North Pacific Ocean. J. Oceanogr. 72, 67–76.

Bennett, B.G., 2002. Worldwide dispersion adn deposition of radionuclides produced in atmospheric tests. Health Phys. 82, 644–655.

Bezhenar, R., Takata, H., de With, G., Maderich, V., 2021. Planned release of contaminated water from the Fukushima storage tanks into the ocean: Simulation scenarios of radiological impact for aquatic biota and human from seafood consumption. Mar. Pollut. Bull. 173, 112969.

Buesseler, K.O., 2020. Opening the floodgates at Fukushima. Science 369, 621–622.

Chino, M., Nakayama, H., Haruyasu, N., Terada, H., Katata, G., Yamazawa, H., 2011. Preliminary Estimation of Release Amounts of $^{131}$I and $^{137}$Cs Accidentally Discharged from the Fukushima Daiichi Nuclear Power Plant into the Atmosphere. J. Nucl. Sci. Technol. 48, 1129–1134.

CNNC, 2024a. Fuqing Nuclear Power Nuclear Safety Information Disclosure Annual Report 2023. China National Nuclear Corporation. In Chinese. https://www.cnnp.com.cn/cnnp/cydwzd62/fjfqhdyxgs/hdchaqxx11/1415089/2024032722262481131.pdf. (Accessed 17 July 2024).

CNNC, 2024b. Sanmen Nuclear Power Nuclear Safety Information Disclosure Quarterly Report 2023. China National Nuclear Corporation. In Chinese https://www.cnnp.com.cn/cnnp/resource/cms/article/1086769/1396493/2024011816580133727.pdf. (Accessed 17 July 2024).

Eyrolle, F., Ducros, L., Le Dizès, S., Beaugelin-Seiller, K., Charmasson, S., Boyer, P., Cossonnet, C., 2018. An updated review on tritium in the environment. J. Environ. Radioact. 181, 128–137.

Ferreira, M.F., Turner, A., Vernon, E.L., Grisolia, C., Lebaron-Jacobs, L., Malard, V., Jha, A.N., 2023. Tritium: Its relevance, sources and impacts on non-human biota. Sci. Total Environ. 876, 162816.

Fukushima Prefecture, 2023. Result of Sea Area Monitoring Associated with the Discharge of ALPS Treated Water into the Sea. In Japanese. https://www.pref.fukushima.lg.jp/site/portal/genan208.html. (Accessed 10 December 2023).

Haynes, W.M. (Ed.), 2017. CRC Handbook of Chemistry and Physics. 97th edition. CRC Press, Boca Raton.
IAEA, 2005. Worldwide marine radioactivity studies (WOMARS): Radionuclide levels in oceans and seas. International Atomic Energy Agency, Vienna.
IAEA, 2023. IAEA Comprehensive Report on the Safety Review of the ALPS-treated water at the Fukushima Daiichi Nuclear Power Station. International Atomic Energy Agency, Vienna.
ICRP 2002. Basic Anatomical and Physiological Data for Use in Radiological Protection Reference Values. ICRP Publication 89. Annals of the ICRP, vol. 32.
Jean-Baptiste, P., Mantisi, F., Dapoigny, A., Stievenard, M., 1992. Design and performance of a mass spectrometric facility for measuring helium isotopes in natural waters and for low-level tritium determination by the $^3$He ingrowth method. Int. J. Radiat. Appl. Instrumentation. Part 43, 881–891.
Kaizer, J., Kumamoto, Y., Molnár, M., Palcsu, L., Povinec, P.P., 2020. Temporal changes in tritium and radiocarbon concentrations in the western North Pacific Ocean (1993–2012). J. Environ. Radioact. 218, 106238.
KEP, 2024. Regarding "tritium" generated during the operation of the Sendai Nuclear Power Plant. Kyushu Electric Power Company. In Japanese. https://www.kyuden.co.jp/var/rev0/0532/6037/n9az8syD.pdf. (Accessed 17 July 2024).
KHNP, 2024. Environmental radioactivity survey and evaluation report around nuclear power plants 2023. Korea Hydro & Nuclear Power Company. In Korean. https://npp.khnp.co.kr/board/view.khnp?boardId=BBS_0000032&menuCd=DOM_000000104003000000&startPage=1&dataSid=10403. (Accessed 17 July 2024).
Kokubon, Y., Fujita, H., Nakano, M., Sumiya, S., 2011. Tritium concentration and diffusion in seawater discharged from Tokai Reprocessing Plant. Prog. Nucl. Sci. Technol. 1, 384–387.
Lehto, J., Koivula, R., Leinonen, H., Tusa, E., Harjula, R., 2019. Removal of radionuclides from Fukushima Daiichi waste effluents. Sep. Purif. Rev. 48, 122–142.
Machida, M., Iwata, A., Yamada, S., Otosaka, S., Kobayashi, T., Funasaka, H., Morita, T., 2023. Estimation of temporal variation of tritium inventory discharged from the port of Fukushima Dai-ichi Nuclear Power Plant:analysis of the temporal variation and comparison with released tritium inventories from Japan and world major nuclear facilities. J. Nucl. Sci. Technol. 60, 258–276.
Maderich, V., Tsumune, D., Bezhenar, R., de With, G., 2024. A critical review and update of modelling of treated water discharging from Fukushima Daiichi NPP. Mar. Pollut. Bull. 198, 115901.
Matsumoto, H., Shimada, Y., Nakamura, A.J., Usami, N., Ojima, M., Kakinuma, S., Shimada, M., Sunaoshi, M., Hirayama, R., Tauchi, H., 2021. Health effects triggered by tritium: How do we get public understanding based on scientifically supported evidence? J. Radiat. Res. 62, 557–563.
METI, 2021. Basic policy on handling of the ALPS treated water. Ministry of Economy, Trade and Industry. https://www.meti.go.jp/english/earthquake/nuclear/decommissioning/pdf/202104_bp_breifing.pdf. (Accessed 10 December 2023).
Ministry of the Environment, 2023. ALPS Treated Water Marine Monitoring Information. https://shorisui-monitoring.env.go.jp/en/. (Accessed 10 December 2023).
Monetti, M.A., 1996. Worldwide Deposition of Strontium-90 Through 1990. USDOE Environmental Measurements Lab., New York, NY (United States).
Nakamura, S., Awata, T., Kiyokawa, H., Ito, H., Ohmura, R., 2023. Tritiated water removal method based on hydrate formation using heavy water as co-precipitant. Chem. Eng. J. 465, 142979.
NRA, 2021. Matters to Determine the Amount of Isotopes That Emit Radiation. National Regulation Authority. In Japanese. https://www.nsr.go.jp/data/000045581.pdf. (Accessed 10 January 2024).

NRA, 2023a. Readings of Sea Area Monitoring near Fukushima Dai-ichi NPP. Nuclear Regulation Authority. https://radioactivity.nra.go.jp/en/contents/17000/16700/24/464_5_20231114.pdf. (Accessed 10 December 2023).

NRA, 2023b. Readings of Sea Area Monitoring at offshore of Miyagi, Fukushima, Ibaraki and Chiba Prefecture (Seawater). Nuclear Regulatory Authority. https://radioactivity.nra.go.jp/ja/results/sea/off-shore (Accessed 18 July 2024).

Palcsu, L., Major, Z., Köllő, Z., Papp, L., 2010. Using an ultrapure $^4$He spike in tritium measurements of environmental water samples by the $^3$He-ingrowth method. Rapid Commun. Mass Spectrom. 24, 698–704.

Povinec, P.P., Aoyama, M., Biddulph, D., Breier, R., Buesseler, K., Chang, C.C., Golser, R., Hou, X.L., Ješkovský, M., Jull, A.J.T., Kaizer, J., Nakano, M., Nies, H., Palcsu, L., Papp, L., Pham, M.K., Steier, P., Zhang, L.Y., 2013. Cesium, iodine and tritium in NW Pacific waters-a comparison of the Fukushima impact with global fallout. Biogeosciences 10, 5481–5496.

Povinec, P.P., Hirose, K., Aoyama, M., Tateda, Y., 2021. Fukushima Accident: 10 years after. Elsevier, New York.

Povinec, P.P., Hirose, K., Honda, T., Ito, T., Scott, E.M.M., Togawa, O., 2004. Spatial distribution of $^3$H, $^{90}$Sr, $^{137}$Cs and $^{239,240}$Pu in surface waters of the Pacific and Indian Oceans - GLOMARD database. J. Environ. Radioact. 76, 113–137.

Povinec, P.P., Lee, S.H., Kwong, L.L.W., Oregioni, B., Jull, A.J.T., Kieser, W.E., Morgenstern, U., Top, Z., 2010. Tritium, radiocarbon, $^{90}$Sr and $^{129}$I in the Pacific and Indian Oceans. Nucl. Instruments Methods Phys. Res. Sect. B Beam Interact. with Mater. Atoms 268, 1214–1218.

Povinec, P.P., Liong Wee Kwong, L., Kaizer, J., Molnár, M., Nies, H., Palcsu, L., Papp, L., Pham, M.K., Jean-Baptiste, P., 2017. Impact of the Fukushima accident on tritium, radiocarbon and radiocesium levels in seawater of the western North Pacific Ocean: A comparison with pre-Fukushima situation. J. Environ. Radioact. 166, 56–66.

Rethinasabapathy, M., Ghoreishian, S.M., Hwang, S., Han, Y., Roh, C., Huh, Y.S., 2023. Recent Progress in Functional Nanomaterials towards the Storage, Separation, and Removal of Tritium. Adv. Mater. 35, 2301589.

Sakuma, K., Machida, M., Kurikami, H., Iwata, A., Yamada, S., Iijima, K., 2022. A modeling approach to estimate $^3$H discharge from rivers: Comparison of discharge from the Fukushima Dai-ichi and inventory in seawater in the Fukushima coastal region. Sci. Total Environ. 806, 151344.

Shozugawa, K., Hori, M., Johnson, T.E., Takahata, N., Sano, Y., Kavasi, N., Sahoo, S.K., Matsuo, M., 2020. Landside tritium leakage over through years from Fukushima Dai-ichi nuclear plant and relationship between countermeasures and contaminated water. Sci. Rep. 10, 19925.

Steinhauser, G., Niisoe, T., Harada, K.H., Shozugawa, K., Schneider, S., Synal, H.A., Walther, C., Christl, M., Nanba, K., Ishikawa, H., Koizumi, A., 2015. Post-accident sporadic releases of airborne radionuclides from the Fukushima Daiichi Nuclear Power Plant site. Environ. Sci. Technol. 49, 14028–14035.

Tatebe, H., Yasuda, I., 2004. Oyashio southward intrusion and cross-gyre transport related to diapycnal upwelling in the Okhotsk Sea. J. Phys. Oceanogr. 34, 2327–2341.

TEPCO, 2023a. ALPS treated water, etc. Conditions. Tokyo Electric Power Company. https://www.tepco.co.jp/en/decommission/progress/watertreatment/alpsstate/index-e.html. (Accessed 20 August 2023).

TEPCO, 2023b. FY2023 Discharge Plan. Tokyo Electric Power Company. https://www.tepco.co.jp/decommission/progress/watertreatment/_assets/images/en/dischargefacility/dscharge_plan-e.pdf. (Accessed 10 December 2023).

TEPCO, 2023c. Results of Radioactive Material Analysis in the Vicinity of the Fukushima Daiichi Nuclear Power Station. Tokyo Electric Power Company. https://www.tepco.co.jp/en/hd/decommission/data/analysis/index-e.html. (Accessed 10 December 2023).

TEPCO, 2023d. Radiological Environmental Impact Assessment Report Regarding the Discharge of ALPS Treated Water into the Sea (Construction stage / Revised version). Tokyo Electric Power Company. https://www.tepco.co.jp/en/hd/newsroom/press/archives/2023/pdf/230220e0101.pdf. (Accessed 24 June 2024).

TEPCO, 2024. Tritium concentration after dilution. Tokyo Electric Power Company. https://www.tepco.co.jp/en/nu/fukushima-np/f1-rt/html-e/f1-alps_fd-month-sel-e.html. (Accessed 10 January 2024).

Tsumune, D., Tsubono, T., Aoyama, M., Hirose, K., 2012. Distribution of oceanic $^{137}$Cs from the Fukushima Dai-ichi Nuclear Power Plant simulated numerically by a regional ocean model. J. Environ. Radioact. 111, 100–108.

Tsunogai, S., Watanabe, S., Honda, M., Aramaki, T., 1995. North Pacific Intermediate Water studied chiefly with radiocarbon. J. Oceanogr. 51, 519–536.

UNSCEAR, 2000. Sources and Effects of Ionizing Radiation, Exposures to the Public from Man-made Sources of Radiation. Annex C, United Nations Scientific Committee on the Effects of Atomic Radiation. United Nations, Ney York, USA.

UNSCEAR, 2008. Sources and Effects of Ionizing Radiation. Report of the United Nations Scientific Committee on the Effects of Atomic Radiation to the General Assembly, United Nation. New York.

UNSCEAR, 2016. Sources, Effects and Risks of ionizing radiation. In: Annex C, Biological Effects of Selected Internal Emitters—tritium, United Nations Scientific Committee on the Effects of Atomic Radiation. United Nations, Ney York, USA.

Watanabe, Y.W., Watanabe, S., Tsunogai, S., 1991. Tritium in the northwestern North Pacific. J. Oceanogr. Soc. Japan 47, 80–93.

WHO, 2017. Guidelines for Drinking-water Quality: Fourth Edition Incorporating the First Addendum. World Health Organization, Geneva.

Zhao, C., Wang, Gang, Zhang, M., Wang, Guansuo, de With, G., Bezhenar, R., Maderich, V., Xia, C., Zhao, B., Jung, K.T., Periáñez, R., Akhir, M.F., Sangmanee, C., Qiao, F., 2021. Transport and dispersion of tritium from the radioactive water of the Fukushima Daiichi nuclear plant. Mar. Poll. Bull. 169, 112515.

**Supplementary data - Appendix A**

Tritium surface activity concentrations (depth of 7-10 m) in the region offshore Fukushima during the Ka'imikai-o-Kanaloa expedition in June 2011

Source(s):
1. Povinec, P.P., Liong Wee Kwong, L., Kaizer, J., Molnár, M., Nies, H., Palcsu, L., Papp, L., Pham, M.K., Jean-Baptiste, P., 2017. Impact of the Fukushima accident on tritium, radiocarbon and radiocesium levels in seawater of the western North Pacific Ocean: A comparison with pre-Fukushima situation. J. Environ. Radioact. 166, 56–66.

| Station | Latitude [°N] | Longitude [°E] | Sampling date | $^3$H activity concentration [Bq/L]* |
|---|---|---|---|---|
| 20 | 37.999 | 143.000 | 13/06/2011 | 0.12 |
| 26 | 36.998 | 141.400 | 15/06/2011 | 0.089 |
| 27 | 36.499 | 141.401 | 15/06/2011 | 0.11 |
| 28 | 36.006 | 141.390 | 16/06/2011 | 0.075 |
| 29 | 36.491 | 142.079 | 16/06/2011 | 0.094 |
| 30 | 36.499 | 142.495 | 17/06/2011 | 0.076 |
| 31 | 36.999 | 142.501 | 17/06/2011 | 0.084 |
| 32 | 36.998 | 142.001 | 18/06/2011 | 0.082 |

* Calculated from the original data that were reported in Tritium Units (TU) where 1 TU is equivalent to the $^3$H activity concentration of 0.119 Bq/L.

Tritium surface activity concentrations (depth of 0.5 m) in the Fukushima coastal region from April to September 2023 as determined by the Marine Ecology Research Institute

Source(s):
1. NRA, 2021. Matters to Determine the Amount of Isotopes That Emit Radiation. National Regulation Authority. In Japanese. https://www.nsr.go.jp/data/000045581.pdf (Accessed 10 January 2024).

| Station | Latitude [°N] | Longitude [°E] | Sampling date | $^3$H activity concentration [Bq/L] |
|---|---|---|---|---|
| M-101 | 37.427 | 141.043 | 21/04/2023 | 0.12 |
| | | | 20/05/2023 | 0.058 |
| | | | 09/06/2023 | 0.082 |
| | | | 07/07/2023 | 0.11 |
| | | | 04/08/2023 | 0.07 |
| | | | 01/09/2023 | 0.066 |
| M-102 | 37.418 | 141.043 | 21/04/2023 | 0.11 |
| | | | 20/05/2023 | 0.098 |
| | | | 09/06/2023 | 0.054 |
| | | | 07/07/2023 | 0.13 |
| | | | 04/08/2023 | 0.064 |
| | | | 01/09/2023 | < 0.052 |
| M-103 | 37.445 | 141.047 | 21/04/2023 | 0.087 |
| | | | 20/05/2023 | 0.094 |
| | | | 09/06/2023 | 0.052 |
| | | | 07/07/2023 | 0.087 |
| | | | 04/08/2023 | 0.079 |
| | | | 01/09/2023 | 0.097 |
| M-104 | 37.402 | 141.047 | 21/04/2023 | 0.056 |
| | | | 20/05/2023 | 0.071 |
| | | | 09/06/2023 | 0.062 |
| | | | 07/07/2023 | 0.11 |
| | | | 04/08/2023 | 0.051 |
| | | | 01/09/2023 | 0.079 |

Tritium surface activity concentrations in the Fukushima coastal region in August and September 2023 as determined by the Ministry of the Environment

Source(s):
1. Ministry of the Environment, 2023. ALPS Treated Water Marine Monitoring Information. https://shorisui-monitoring.env.go.jp/en/. (Accessed 10 December 2023).

| Station | Latitude [°N] | Longitude [°E] | Sampling date | $^3$H activity concentration [Bq/L] |
|---|---|---|---|---|
| E-1 | 37.441 | 141.040 | 25/08/2023 | 0.062 |
|  |  |  | 13/10/2023 | 0.76 |
|  |  |  | 01/11/2023 | 0.12 |
|  |  |  | 14/11/2023 | 3.5 |
| E-2 | 37.434 | 141.039 | 13/09/2023 | 0.085 |
|  |  |  | 14/11/2023 | 0.65 |
| E-3 | 37.433 | 141.046 | 25/08/2023 | 0.13 |
|  |  |  | 14/11/2023 | 0.24 |
| E-4 | 37.429 | 141.051 | 25/08/2023 | 0.71 |
|  |  |  | 13/10/2023 | 0.22 |
|  |  |  | 01/11/2023 | 0.13 |
|  |  |  | 14/11/2023 | 0.22 |
| E-5 | 37.426 | 141.046 | 25/08/2023 | 5.0 |
|  |  |  | 14/11/2023 | 0.2 |
| E-6 | 37.420 | 141.042 | 13/09/2023 | 0.097 |
|  |  |  | 14/11/2023 | 0.23 |

Tritium surface activity concentrations in the Fukushima coastal region in September and October 2023 as determined by the Fukushima Prefecture

Source(s):
1. Fukushima Prefecture, 2023. Result of Sea Area Monitoring Associated with the Discharge of ALPS Treated Water into the Sea. In Japanese. https://www.pref.fukushima.lg.jp/site/portal/genan208.html. (Accessed 10 December 2023).

| Station | Latitude [°N] | Longitude [°E] | Sampling date | $^3$H activity concentration [Bq/L] |
|---|---|---|---|---|
| FP-4 | 37.424 | 141.057 | 03/09/2023 | 0.15 |
| | | | 12/10/2023 | 0.05 |
| | | | 09/11/2023 | 0.17 |
| FP-8 | 37.437 | 141.047 | 03/09/2023 | 0.08 |
| | | | 12/10/2023 | 0.05 |
| | | | 09/11/2023 | 0.28 |
| FP-9 | 37.420 | 141.047 | 03/09/2023 | 0.12 |
| | | | 12/10/2023 | 0.27 |
| | | | 09/11/2023 | 1.6 |

Tritium surface activity concentrations (depth of 0.5 m) in the Fukushima coastal region from August to November 2023 as determined by the Tokyo Electric Power Company

Source(s):
1. TEPCO, 2023. Results of Radioactive Material Analysis in the Vicinity of the Fukushima Daiichi Nuclear Power Station. Tokyo Electric Power Company. https://www.tepco.co.jp/en/hd/decommission/data/analysis/index-e.html. (Accessed 10 December 2023).

| Station | Latitude [°N] | Longitude [°E] | Sampling date | $^3$H activity concentration [Bq/L] |
|---|---|---|---|---|
| T-A1 | 37.441 | 141.051 | 21/08/2023 | < 0.36 |
| | | | 24/08/2023 | < 0.32 |
| | | | 26/08/2023 | 0.13 |
| | | | 30/08/2023 | 1.1 |
| | | | 04/09/2023 | 1.1 |
| | | | 25/09/2023 | < 0.3 |
| | | | 05/10/2023 | 0.088 |
| | | | 16/10/2023 | 0.50 |
| | | | 23/10/2023 | 0.37 |
| | | | 01/11/2023 | < 0.31 |
| | | | 02/11/2023 | < 0.31 |
| | | | 06/11/2023 | < 0.39 |
| | | | 13/11/2023 | 0.082 |
| | | | 20/11/2023 | 1.5 |
| | | | 27/11/2023 | < 0.36 |
| T-0-1 | 37.431 | 141.040 | 21/08/2023 | < 0.35 |
| | | | 24/08/2023 | < 0.34 |
| | | | 26/08/2023 | 0.66 |
| | | | 30/08/2023 | < 0.32 |
| | | | 04/09/2023 | < 0.34 |
| | | | 11/09/2023 | 0.10 |
| | | | 25/09/2023 | < 0.35 |
| | | | 05/10/2023 | < 0.31 |
| | | | 09/10/2023 | 1.4 |
| | | | 16/10/2023 | 1.0 |
| | | | 23/10/2023 | 1.3 |
| | | | 01/11/2023 | < 0.35 |
| | | | 02/11/2023 | < 0.36 |
| | | | 06/11/2023 | 0.36 |
| | | | 13/11/2023 | 0.15 |
| | | | 20/11/2023 | 1.2 |
| | | | 27/11/2023 | 0.38 |
| T-0-1A | 37.431 | 141.047 | 21/08/2023 | < 0.34 |
| | | | 24/08/2023 | 2.6 |
| | | | 26/08/2023 | 0.087 |
| | | | 30/08/2023 | 0.43 |
| | | | 04/09/2023 | < 0.33 |
| | | | 11/09/2023 | 0.12 |
| | | | 25/09/2023 | < 0.35 |
| | | | 05/10/2023 | 5.2 |
| | | | 09/10/2023 | 12 |
| | | | 16/10/2023 | 14 |
| | | | 23/10/2023 | 0.71 |
| | | | 01/11/2023 | < 0.34 |
| | | | 02/11/2023 | 6.9 |
| | | | 06/11/2023 | 9.5 |
| | | | 13/11/2023 | 0.14 |
| | | | 20/11/2023 | 1.0 |
| | | | 27/11/2023 | < 0.33 |
| T-A2 | 37.425 | 141.051 | 21/08/2023 | < 0.36 |
| | | | 24/08/2023 | < 0.32 |
| | | | 26/08/2023 | 0.065 |
| | | | 30/08/2023 | 1.5 |
| | | | 04/09/2023 | 0.88 |
| | | | 25/09/2023 | < 0.30 |
| | | | 05/10/2023 | < 0.064 |
| | | | 16/10/2023 | 0.56 |
| | | | 23/10/2023 | < 0.31 |
| | | | 01/11/2023 | < 0.31 |
| | | | 02/11/2023 | < 0.30 |
| | | | 06/11/2023 | < 0.38 |
| | | | 13/11/2023 | 0.16 |
| | | | 20/11/2023 | 0.60 |
| | | | 27/11/2023 | < 0.36 |
| T-0-2 | 37.423 | 141.047 | 21/08/2023 | < 0.32 |
| | | | 24/08/2023 | < 0.35 |
| | | | 26/08/2023 | 0.92 |
| | | | 30/08/2023 | 1.4 |
| | | | 04/09/2023 | 0.74 |
| | | | 25/09/2023 | < 0.30 |
| | | | 05/10/2023 | < 0.33 |
| | | | 02/11/2023 | < 0.37 |
| | | | 20/11/2023 | 0.77 |
| | | | 27/11/2023 | < 0.26 |
| T-0-3 | 37.416 | 141.040 | 21/08/2023 | < 0.33 |
| | | | 24/08/2023 | < 0.34 |
| | | | 26/08/2023 | 0.14 |
| | | | 30/08/2023 | < 0.31 |
| | | | 04/09/2023 | < 0.34 |
| | | | 11/09/2023 | 0.16 |
| | | | 25/09/2023 | < 0.35 |
| | | | 05/10/2023 | < 0.32 |
| | | | 09/10/2023 | 0.45 |
| | | | 16/10/2023 | 1.0 |
| | | | 23/10/2023 | 1.0 |
| | | | 01/11/2023 | < 0.34 |
| | | | 02/11/2023 | < 0.36 |
| | | | 06/11/2023 | < 0.31 |
| | | | 13/11/2023 | 0.44 |
| | | | 20/11/2023 | 0.92 |
| | | | 27/11/2023 | < 0.33 |
| T-0-3A | 37.416 | 141.040 | 21/08/2023 | < 0.33 |
| | | | 24/08/2023 | < 0.33 |
| | | | 26/08/2023 | < 0.068 |
| | | | 30/08/2023 | < 0.32 |
| | | | 04/09/2023 | < 0.33 |
| | | | 11/09/2023 | 0.10 |
| | | | 25/09/2023 | < 0.35 |
| | | | 05/10/2023 | < 0.32 |
| | | | 09/10/2023 | < 0.072 |
| | | | 16/10/2023 | 0.74 |
| | | | 23/10/2023 | < 0.33 |
| | | | 01/11/2023 | < 0.32 |
| | | | 02/11/2023 | < 0.26 |
| | | | 06/11/2023 | 0.54 |
| | | | 13/11/2023 | 0.49 |
| | | | 20/11/2023 | 0.87 |
| | | | 27/11/2023 | < 0.33 |
| T-A3 | 37.409 | 141.051 | 21/08/2023 | < 0.36 |
| | | | 24/08/2023 | < 0.32 |
| | | | 26/08/2023 | < 0.072 |
| | | | 30/08/2023 | 1.1 |
| | | | 04/09/2023 | 0.82 |
| | | | 25/09/2023 | < 0.29 |
| | | | 05/10/2023 | < 0.077 |
| | | | 16/10/2023 | 0.8 |
| | | | 23/10/2023 | < 0.32 |
| | | | 01/11/2023 | < 0.32 |
| | | | 02/11/2023 | < 0.31 |
| | | | 06/11/2023 | < 0.39 |
| | | | 13/11/2023 | 0.15 |
| | | | 20/11/2023 | 0.37 |
| | | | 27/11/2023 | < 0.36 |